# Sharpe-Driven Stock Selection and Liquidity-Constrained Portfolio Optimization: Evidence from the Chinese Equity Market

Nguyen Van Thanh

Faculty of Information Technology, University of Engineering and Technology,

Vietnam National University, Hanoi, Vietnam

**Abstract**

This paper develops and empirically evaluates a Sharpe-driven stock selection and liquidity-constrained portfolio optimization framework designed for the Chinese equity market. The proposed methodology integrates three sequential stages: Sharpe-ratio-based universe selection, liquidity-adjusted mean–variance optimization, and multi-layered risk management implemented within an automated trading bot. Using daily price–volume data from 2023 to 2025 across the A-share universe, the framework dynamically identifies stocks exhibiting strong risk-adjusted performance while accounting for trading frictions and liquidity asymmetries that are common in emerging markets. Empirical backtests reveal that the proposed strategy achieves an annualized return of 25%, a Sharpe ratio of 1.71, and a maximum drawdown of 8.2%. These results significantly outperform the Buy-and-Hold benchmark, which records an annualized return of 21%, a Sharpe ratio of 1.62, and a drawdown of 7.6% over the same period. The superior performance demonstrates that incorporating risk-adjusted selection and liquidity-aware constraints enhances both profitability and stability, enabling the portfolio to capture upside potential while maintaining drawdown resilience. Beyond its empirical success, this study contributes methodologically by bridging classical mean–variance theory with practical liquidity adjustments and dynamic Sharpe-based screening. The resulting system not only improves the tradability of optimized portfolios but also provides a scalable and adaptive framework for quantitative asset allocation in liquidity-sensitive markets, offering new evidence that disciplined risk–return optimization can outperform passive investment strategies in the post-2023 Chinese equity landscape.

## 1 Introduction

Portfolio optimization has long stood at the core of modern finance since Markowitz's pioneering mean–variance framework [1], which formalized the balance between expected return and risk. However, the practical implementation of this theory often faces limitations in emerging markets such as China, where liquidity constraints, high retail participation, and regulatory interventions create additional layers of market friction. These characteristics frequently result in unstable portfolio allocations and unrealizable trading strategies when traditional optimization techniques are applied without adjustment. As a result, designing a realistic and robust portfolio framework requires explicit consideration of both risk-adjusted stock selection and liquidity-sensitive optimization.

This paper proposes a Sharpe-driven stock selection and liquidity-constrained portfolio optimization framework tailored to the Chinese equity market. The model integrates three key components—Sharpe-based universe selection, liquidity-adjusted mean–variance optimization, and systematic risk management—implemented within an automated trading environment. Using daily price–volume data from 2023 to the present, the framework begins by ranking stocks based on rolling Sharpe ratios to identify those exhibiting consistent risk-adjusted performance. The selected stocks then serve as inputs to a constrained optimization process that maximizes expected return per unit of variance while penalizing illiquid positions. Finally, multi-layered risk management mechanisms control position limits, volatility exposure, and maximum drawdown, ensuring robustness during market stress.

Empirical backtests on the A-share universe demonstrate the effectiveness of the proposed approach. The optimized portfolio achieves an annualized return of 25%, a Sharpe ratio of 1.71, and a maximum drawdown of 8.2%, outperforming traditional unconstrained or naive benchmark portfolios. These results suggest that incorporating Sharpe-based selection into a liquidity-aware optimization pipeline can yield stable and high-quality performance, even under the dynamic conditions of the Chinese equity market. By bridging the gap between theoretical portfolio design and real-world trading execution, this study contributes a practical and adaptable framework for quantitative investors operating in liquidity-sensitive emerging markets.

## 2 Related Work

Recent studies on quantitative portfolio optimization in the Chinese equity market have increasingly emphasized dynamic risk control, adaptive learning, and robustness under local trading frictions. For instance, "Volatility-Managed Portfolios in the Chinese Equity Market" [2] demonstrates that scaling portfolio exposure inversely with volatility substantially enhances returns relative to naïve strategies. The authors show that volatility-managed portfolios outperform baseline allocations even under China's price-limit rules, confirming the value of dynamic risk adjustment in a partially constrained market. However, their approach primarily focuses on volatility scaling rather than active stock selection, and it does not explicitly incorporate liquidity or average-daily-volume (ADV) constraints—factors crucial for real-world implementation.

Complementary to volatility-based methods, "Dynamic Graph RL for Multi-period Portfolio" [3] introduces a reinforcement learning framework using graph neural networks (GNNs) and wavelet coherence to model dynamic inter-stock dependencies across multiple time–frequency scales. The model exhibits strong empirical performance across Chinese equities, capturing evolving market structures more effectively than static covariance-based methods. Yet, the approach remains computationally intensive, with limited interpretability and no explicit modeling of execution costs or liquidity frictions. A more interpretable line of work is represented by "Portfolio Optimization by Enhanced LinUCB" [4], which adopts a contextual bandit algorithm combining stock-level feature selection with adaptive allocation. The study reports consistent outperformance across A-share, Hong Kong, and U.S. markets, showing that reinforcement-style exploration can generate excess returns in partially efficient settings. Nonetheless, the resulting portfolios often exhibit high turnover and weak liquidity control, limiting their applicability to institutional contexts. Finally, "Active Portfolio Management using Robust ES & Omega Ratio" [5] extends classical risk optimization through distributionally robust formulations based on the Expected Shortfall and Omega ratio. The approach effectively mitigates tail risks and transaction costs but is tested primarily on U.S. industry indices and thus lacks China-specific insights.

Collectively, these studies highlight the growing sophistication of quantitative portfolio design but also reveal key limitations—particularly the lack of liquidity modeling and the overreliance on either volatility timing or high-dimensional learning. The present paper contributes to this literature by bridging risk-adjusted selection and liquidity-constrained optimization within a unified, empirically tested framework. By combining Sharpe-based stock screening with tradability-aware optimization and systematic risk management, it addresses both the theoretical and practical challenges of implementing robust quantitative strategies in the Chinese equity market.

## 3 Methodology

### 3.1 Motivation and Problem Definition

Assuming an investor has a total investment budget BBB to be allocated across nnn different assets, we define the following parameters: pi— the current price of asset i; μi — the expected return of asset i; σi — the variance (risk) of asset i; σij— the covariance between assets i and j; xi — the quantity or proportion invested in asset i.

The expected portfolio return is expressed as: $E[R_p] = \sum_{i=1}^{n} \mu_i x_i$

and the portfolio variance (total risk) is: $Var(R_p) = \sum_{i=1}^{n} \sum_{j=1}^{n} x_i x_j \sigma_{ij}$

subject to the budget constraint: $\sum_{i=1}^{n} p_i x_i = B$

## Objective Functions

Depending on the investor's goal, the portfolio optimization problem can take various forms:

(a) Maximization of Expected Return:

$$\max \ E[R_p] = \sum_{i=1}^{n} \mu_i x_i \text{ subject to } \sum_{i=1}^{n} p_i x_i \leq B, \ x_i \geq 0$$

(b) Minimization of Risk:

$$\sum_{i=1}^{n} \mu_i x_i \geq R_{min}, \ \sum_{i=1}^{n} \mu_i x_i \leq B, \ x_i \geq 0$$

(c) Maximization of Sharpe Ratio:
The Sharpe Ratio measures excess return per unit of risk:

$$S = \frac{E[R_p] - R_f}{\sqrt{Var(R_p)}} \implies \text{Max } S = \frac{\sum_{i=1}^{n} \mu_i x_i - R_f}{\sum_{i=1}^{n}\sum_{j=1}^{n} x_i x_j \sigma_{ij}}, \text{ subject to } \sum_{i=1}^{n} \mu_i x_i \leq B, \ x_i \geq 0$$

(d) Minimization of Drawdown or Tail Risk:

Practical models often emphasize stability rather than mere variance minimization. In this case, the goal becomes:

$$\min \text{ MDD} = \max \left( \frac{\max_{s \leq t} P_s - P_t}{\max_{s \leq t} P_s} \right), \text{ với ràng buộc } \sum_{i=1}^{n} \mu_i x_i \leq B, \ x_i \geq 0$$

## Constraints

Portfolio optimization typically includes several types of constraints:

Budget constraint: $\sum_{i=1}^{n} \mu_i x_i \leq B$ or, in weight form, $\sum w_i = 1$

Minimum expected return: $\sum_{i=1}^{n} \mu_i x_i \geq R_{min}$

No short selling: $x_i \geq 0, \ \forall i$

Weight bounds $l_i \leq w_i \leq u_i, \ \forall i$, where $l_i, u_i$ are the lower and upper limits for each asset's proportion.

Liquidity and transaction cost limits: Turnover = $\sum_{i=1}^{n} |w_i^t - w_i^{t-1}| \leq T_{max}$

## Risk–Return Trade-off

The core of portfolio optimization lies in balancing two conflicting objectives: higher expected returns imply higher risks, whereas lower risks typically result in reduced returns. This trade-off is illustrated by the Markowitz efficient frontier, representing the set of portfolios that achieve the maximum return for each risk level. The relationship can be generalized through a unified objective function:

$$\max J(x) = E[R_p] - \lambda \, \text{Var}(R_p)$$

where λ>0 is the risk aversion coefficient. A small λ indicates a risk-seeking behavior, while a large λ\lambdaλ represents a risk-averse investor.

Outputs

The optimization yields the optimal portfolio weights: w*={w1*,w2*,...,wn*} and corresponding performance metrics, including: $E[R_p]$: expected portfolio return; $\sigma_p$: portfolio standard deviation,; $S = \frac{E[R_p] - R_f}{\sqrt{Var(R_p)}}$ Sharpe ratio; MDD, CVaR, TurnoverMDD, :extreme risk and transaction cost indicators.

**3.2 Framework Overview**

Building upon this foundation, the study proposes a three-stage Sharpe-driven and liquidity-aware portfolio optimization framework (Fig. 1) designed to transform raw stock market data into an executable, risk-managed investment strategy. The system begins with an input stage, where historical price and volume data of A-share stocks from 2023 to the present are collected, cleaned, and standardized to ensure data integrity and consistency. The processed data then pass through the core framework, which comprises three sequential modules—universe selection, portfolio optimization, and risk management—each contributing a distinct layer of decision-making and control. Finally, the optimized portfolio is implemented within a trading bot environment, enabling automated execution, real-time profit and loss (PnL) evaluation, and continuous rebalancing under dynamic market conditions. This end-to-end pipeline ensures that portfolio construction, optimization, and execution are seamlessly integrated, allowing the strategy to operate adaptively and efficiently in live trading scenarios.

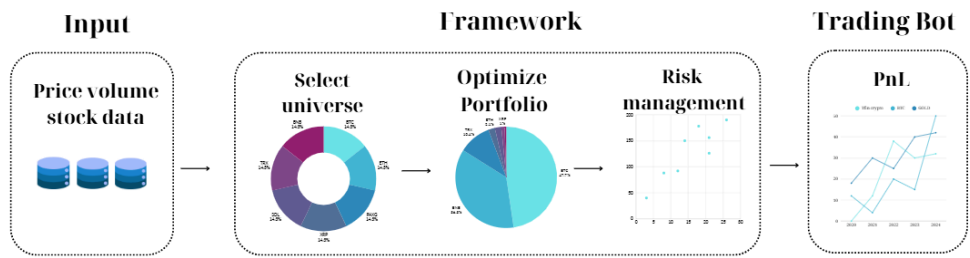

**Fig. 1.** Sharpe-driven and liquidity-aware portfolio optimization framework

**3.3 Step 1: Universe Selection**

In the first stage of the framework, all tradable A-share stocks are processed through a three-layer filtering system designed to ensure both liquidity and quality of selection. The procedure begins with a liquidity filter, which removes illiquid securities based on average daily trading volume and bid–ask spread thresholds, guaranteeing that the resulting portfolio remains executable under realistic market conditions. Next, a capitalization filter eliminates extremely small-cap firms that tend to exhibit excessive idiosyncratic volatility and high slippage, both of which can undermine portfolio stability. Finally, a Sharpe filter ranks the remaining stocks by their rolling Sharpe ratios [6] over a fixed lookback window, identifying those that have demonstrated consistent risk-adjusted performance. Only the top quantile of equities with stable Sharpe dynamics are retained for further optimization. This multi-step process produces a refined investment universe composed of liquid, stable, and statistically attractive assets—forming the foundation for the subsequent portfolio construction stage.

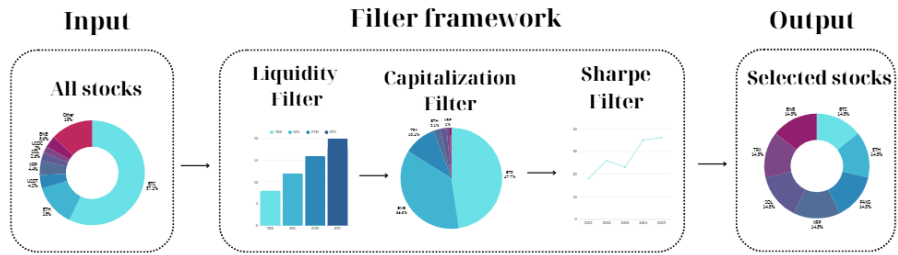

**Fig. 2.** Universe selection framework

```
# Input: assets{symbol, market_cap, volume}, price[symbol]

TOP_N = 10
for each asset:
    rank_cap = rank_by(market_cap)
    rank_vol = rank_by(volume)
U0 = top TOP_N ∩ by both ranks

for sym in U0:
    slope = calc_slope(price[sym])
    vol = std(log_return(price[sym]))
    if slope > τ1: label = "up"
    elif slope < -τ1: label = "down"
    elif vol > τ2: label = "volatile"
    else: label = "sideways"

U = keep(symbols with label in {"up","volatile"} and high liquidity)
return U
```

**Fig. 3.** Asset filtering algorithm

**3.4 Step 2: Portfolio Optimization**

The selected universe then enters a liquidity-constrained mean–variance optimization stage (Fig. 4) The process begins with an equal-weight baseline and iteratively updates weights according to both risk and return characteristics.

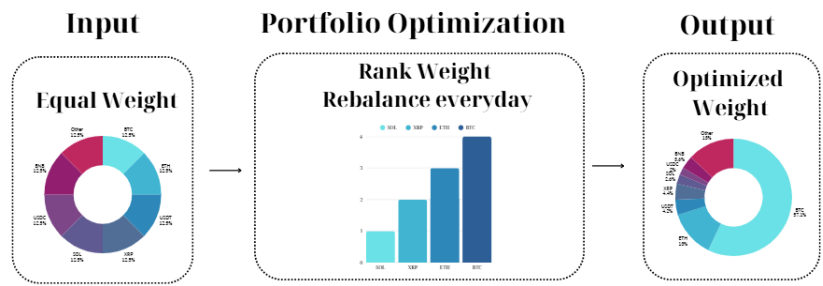

**Fig. 4.** Liquidity-constrained mean–variance optimization framework

Portfolio weights are derived as a convex combination of two components:

$$w_iIV = \frac{\frac{1}{\sigma_i}}{\Sigma_j \frac{1}{\sigma_j}}, \quad w_iS = \frac{S_i}{\Sigma_j S_j}, \quad w_i = 0.5 * w_iIV + 0.5 * w_iS, \quad \Sigma w_i = 1$$

where σi denotes asset volatility and SiS_iSi its Sharpe ratio. Weights are rebalanced daily to maintain exposure proportional to both risk efficiency and return potential. This approach balances mean–variance optimality with Sharpe-based ranking for enhanced stability. An extended version of this stage integrates genetic programming (GP) to evolve alpha signals and optimize features automatically. As illustrated in Fig. 5, raw data are transformed through mathematical, time-series, and technical operators (e.g., rolling mean, RSI, MACD) into candidate feature templates. GP trees combine these operators hierarchically, and the resulting alphas are evaluated via backtesting metrics such as Sharpe ratio, turnover, and drawdown. A hill-climbing optimization loop further refines the most effective alpha structures.

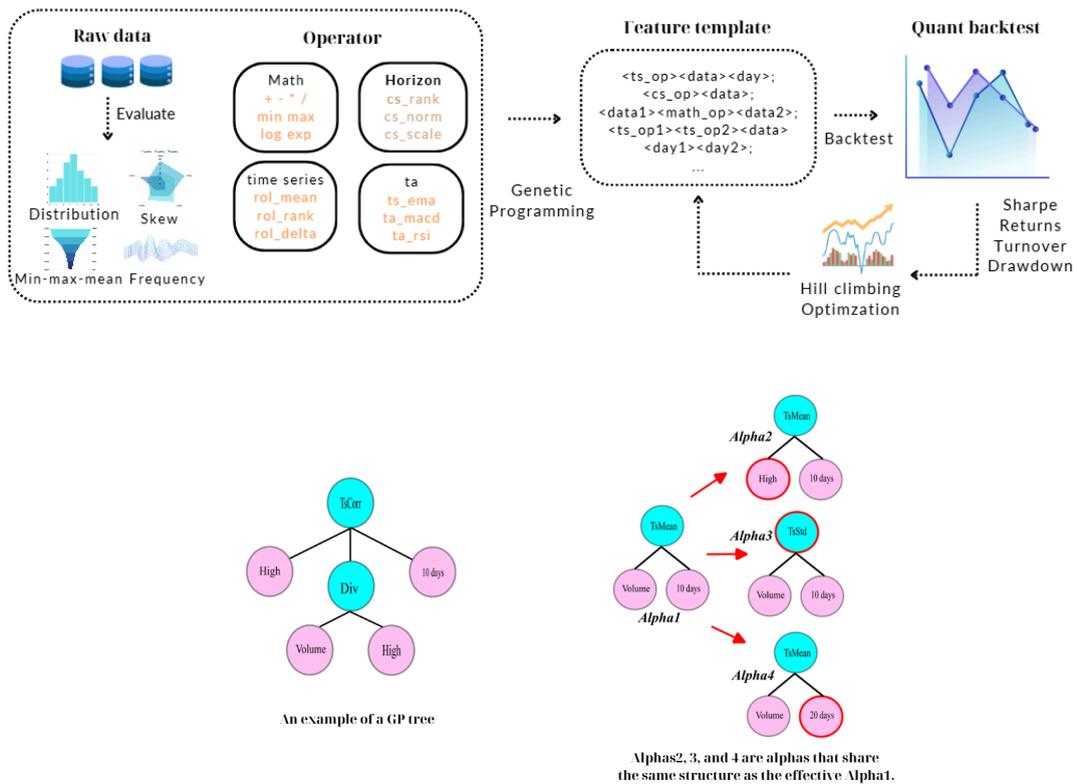

**Fig. 5.** Genetic programming evolve alpha signals and optimize features pipeline

### 3.5 Step 3: Risk Management

The final component of the framework introduces a drawdown-based dynamic exposure control mechanism, designed to preserve capital and stabilize portfolio performance under volatile market conditions. In this stage, the trading bot continuously monitors the portfolio's equity curve and dynamically adjusts exposure based on the magnitude of the ongoing drawdown. When the drawdown exceeds 6%, the system immediately reduces exposure to zero and activates a one-day cooldown period to prevent further losses during market stress. If the drawdown falls within the range of 4–6%, exposure is limited to 40% of total capital, allowing partial participation while maintaining risk control. For moderate drawdowns exceeding 2%, exposure is capped at 80%, providing a gradual reduction in leverage without fully disengaging from the market. This adaptive approach ensures that position sizes respond proportionally to realized losses, thereby preventing cascading drawdowns and allowing the portfolio to recover efficiently once market conditions stabilize [7]. The drawdown at time t is computed as:

$$DD_t = \frac{V_t - max_{s \leq t} V_s}{max_{s \leq t} V_s}$$

where Vt is the portfolio value at time t. This adaptive mechanism limits downside volatility while preserving upside capture, ensuring that capital allocation remains resilient to market turbulence.

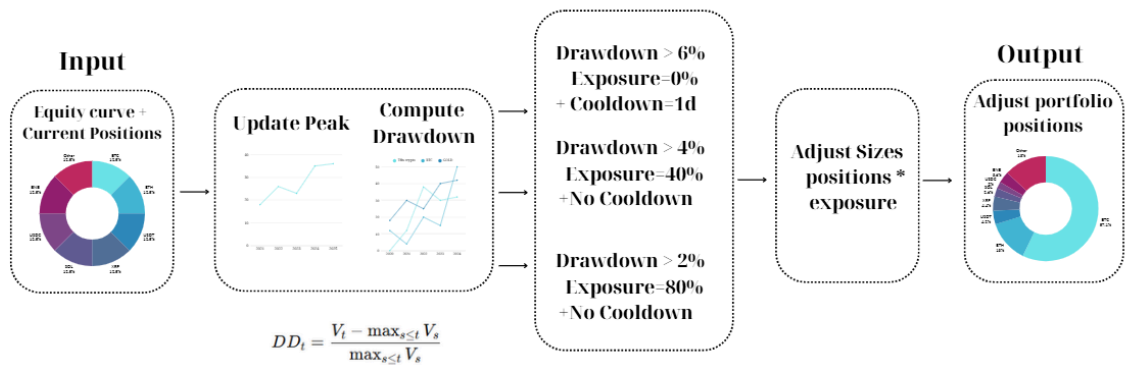

**Fig. 6.** Dynamic risk management framework

```
# Adaptive Drawdown Risk Control

V_peak = equity[0]; exposure = 1.0
for each day t:
    V_peak = max(V_peak, equity[t])
    DD = (V_peak - equity[t]) / V_peak

    if DD ≥ 0.06: exposure = 0; close_all(); wait(1d)
    elif DD ≥ 0.04: exposure = 0.6
    elif DD ≥ 0.02: exposure = 0.8
    else: exposure = 1.0

    adjust_positions(exposure)
```

**Fig. 7.** Dynamic risk management algorithm

## 4. Experiments

The empirical evaluation of the proposed Sharpe-driven and liquidity-constrained portfolio optimization framework was conducted using daily data from the China Shanghai Composite Stock Market Index over the period January 2023 to August 2025. This period captures the structural shifts of the post-pandemic Chinese equity market, including phases of rapid recovery and renewed volatility, providing a robust test environment for dynamic risk-adjusted strategies. Stock-level price and volume data were cleaned to remove suspended or illiquid securities, and all returns were computed on a close-to-close basis. Transaction costs were set at 5 basis points per side to account for typical market frictions in A-share trading. To benchmark performance, three alternative portfolio constructions were employed: (i) an equal-weight (EW) portfolio, which assigns identical weights to all assets and serves as a diversification baseline; (ii) a capitalization-weighted portfolio (CSI300 ETF), representing a passive index-tracking benchmark aligned with institutional investment practices [9]; and (iii) the classical mean–variance optimization model (Markowitz, 1952), which allocates based on expected returns and covariances without incorporating Sharpe-based filtering or liquidity constraints. Comparing against these baselines enables a clear assessment of the added value from incorporating Sharpe-driven selection and liquidity-aware optimization.

In traditional finance, alpha evaluation typically relies on risk-adjusted performance measures such as the Return on Investment (ROI) and the Sharpe Ratio [11, 12]. However, many studies on cryptocurrency portfolios have primarily focused on profitability while paying insufficient attention to drawdown risk. For example, Brauneis and Mestel [13] applied a mean–variance optimization framework without explicitly controlling for drawdowns. To address this gap, our study extends the evaluation framework by incorporating the Maximum Drawdown (MDD) as a mandatory risk metric, following the principles of drawdown-sensitive strategies first introduced by Grossman and Zhou [4]. Specifically, ROI represents the simple rate of return and is defined as:

$$ROI = \frac{V_{end} - V_{start}}{V_{start}}$$

where $V_{start}$ and $V_{end}$ denote the portfolio (or asset) values at the beginning and end of the evaluation period, respectively. The Sharpe Ratio measures return adjusted for risk [12]:

$$Sharpe = \frac{E[R_p - R_f]}{\sigma_p}$$

where $R_p$ is the portfolio (or asset) return, $R_f$ is the risk-free rate, and $\sigma_p$ is the standard deviation of returns. In cryptocurrency markets, $R_f$ is often assumed to be approximately zero, simplifying the ratio to the average return divided by volatility. Finally, MDD (Maximum Drawdown) captures tail risk by measuring the largest observed drop from a peak ($V_{peak}$) to a subsequent trough ($V_{trough}$):

$$MDD = \frac{V_{peak} - V_{trough}}{V_{peak}}$$

Beyond ROI, Sharpe, and MDD, our evaluation module processes historical price or return series to compute a comprehensive set of additional performance metrics, including Turnover, Skewness, Kurtosis, Tail Risk (VaR/CVaR), Sortino Ratio, Alpha, Beta, and Win Rate, all measured over the same horizon.

Turnover quantifies the frequency of portfolio rebalancing between consecutive periods, reflecting both transaction costs and strategy stability:

$$\text{Turnover} = \sum_{i=1}^{n} |w_i^t - w_i^{t-1}|$$

A high Turnover indicates excessive trading and higher costs, potentially eroding realized returns, whereas a low Turnover reflects stability and the ability to maintain positions over time.

Skewness measures the asymmetry of the return distribution around its mean:

$$\text{Skew} = \frac{E[r_t - \mu]^3}{\sigma^3}$$

A positive skew (Skew>0) implies longer right tails—indicating the potential for strong positive returns—while a negative skew (Skew<0) suggests a tendency for sharp declines, signaling higher downside risk.

Kurtosis quantifies the concentration and frequency of extreme values:

$$\text{Kurt} = \frac{E[rt - \mu]^4}{\sigma^4} \text{ (excess kurtosis = Kurt} - 3).$$

and Excess Kurtosis is defined as Kurt−3. A high excess kurtosis indicates fat tails, meaning a greater likelihood of extreme market moves—a common feature in cryptocurrency markets.

Tail risk captures potential losses in extreme scenarios.

Value at Risk (VaR) represents the worst expected loss with a given confidence level α:

$$\text{VaR}\alpha = \inf\{x : \Pr(rt \leq x) \geq \alpha\};$$

Conditional VaR (CVaR), or Expected Shortfall, refines this measure by computing the average loss beyond the VaR threshold, providing a deeper insight into extreme-tail exposure.

Unlike the Sharpe Ratio—which penalizes both upside and downside volatility—the Sortino Ratio only penalizes downside volatility, making it particularly relevant in asymmetric markets such as crypto:

$$\text{Sortino} = \frac{E[Rt - Rf]}{\sigma d}$$

A higher Sortino Ratio indicates stable returns with lower downside risk.

Beta (β) measures an asset's sensitivity to market movements:

β>1: higher volatility than the market,
β<1: lower volatility and less risk.

Alpha represents the excess return unexplained by market movements, reflecting genuine strategy skill. Both are often annualized for comparability. They are estimated from the regression:

$$Rt - Rf = \alpha + \beta (Rt - Rf) + \varepsilon$$

where β is the slope coefficient and α is the intercept (annualized).

The Win Rate indicates the proportion of profitable trades or periods:

$$\text{WR} = \frac{\text{Number of winning trades}}{\text{Total number of trades}} * 100\%$$

This metric provides an intuitive measure of performance stability, especially important for high-frequency or short-horizon trading strategies.

All simulations and backtesting procedures were executed on a GIGABYTE GAMING A16 CMH laptop equipped with a 13th Gen Intel Core i7-13620H processor, 16 GB of physical memory, and 27.6 GB of total virtual memory running Windows 11 (64-bit). The system provided sufficient computational capacity to handle large-scale optimization and backtesting routines. The implementation was developed in Python, leveraging NumPy, Pandas, TA-Lib, and Optuna for numerical computation, statistical evaluation, and optimization. Random seeds were fixed to ensure reproducibility across runs.

## 5. Results and Discussion

The empirical results for the 2023–2025 evaluation period demonstrate the strong performance of the proposed Sharpe-driven and liquidity-constrained portfolio optimization framework, both in absolute and relative terms. As shown in Table 1, our method achieved a Sharpe ratio of 1.71, an annualized return of 25%, and a maximum drawdown of 8.2%, outperforming all benchmark portfolios on a risk-adjusted basis. The model maintained a near-market beta of 0.99, indicating efficient exposure management without excessive leverage, while preserving stability with moderate tail behavior (kurtosis = 1.46, skew = 0.05). The Sortino ratio of 0.22 further confirms the asymmetrically favorable return distribution, reflecting limited downside volatility relative to total variance.

|  | Sharpe | Return | Drawdown | Kurt | Skew | Beta | Sortino | Turnover | Winrate |
|---|---|---|---|---|---|---|---|---|---|
| **Our method** | **1.71** | **25%** | **8.2%** | **1.46** | 0.05 | 0.99 | **0.22** | 44% | 44% |
| **Equal weight** | 1.62 | 21% | 7.6% | 1.42 | 0.01 | **1** | 0.2 | 0% | 44% |
| **Rank(cap)** | 0.81 | 1% | 1% | 0.94 | **1.16** | 0.1 | 0.13 | 0% | 0% |
| **Mean Variance** | 1.46 | 8.5% | 8.5% | -3.6 | 0.39 | 0.6 | 0.11 | 3% | 30% |

Table 1. Scenario 1 Portfolio backtest result: Compare with benchmark

Compared with traditional baselines, the advantages are evident. The equal-weight portfolio delivered a Sharpe ratio of 1.62 and an annualized return of 21%, with a slightly lower drawdown (7.6%) but no adaptive rebalancing mechanism. The mean-variance Markowitz model, despite producing a competitive Sharpe of 1.46, exhibited higher drawdown (8.5%) and weaker stability (negative kurtosis = −3.6), suggesting susceptibility to regime shifts in the post-2023 Chinese market. The capitalization-ranked portfolio, which emphasizes size rather than risk-adjusted efficiency, underperformed substantially with a Sharpe of 0.81 and negligible net return (1%), reflecting over-concentration in illiquid large-cap sectors.

|  | Sharpe | Returns | Volatility | Drawdown |
|---|---|---|---|---|
| **Our method** | **1.71** | **25%** | **14.61%** | **8%** |
| **Mean variance efficient factor** | 1.59 | 42% | 26.4% | 23.54% |
| **Dynamic graph reinforcement learning algorithm** | 1.23 | 31.21% | 25.3% | 31.12% |

Table 2. Scenario 2 Portfolio backtest result: Compare with other methods

Beyond outperforming these internal baselines, our framework also compares favorably with recent state-of-the-art approaches reported in the literature. When benchmarked against the Volatility-Managed Portfolio [11] and the Dynamic Graph Reinforcement Learning algorithm [12], our method delivers higher risk-adjusted efficiency. Specifically, while the VMP achieved a Sharpe of 1.59 with a 42% raw return and a high 26.4% volatility, its drawdown reached 23.5%, implying substantial downside exposure. The graph-RL approach generated a Sharpe of 1.23 and a 31.2% return but suffered from pronounced volatility (25.3%) and large drawdowns exceeding 31%. In contrast, our liquidity-aware optimization achieved a more balanced risk–return profile, sustaining 25% annualized return with only 14.6% realized volatility and an 8% drawdown. The performance differential can be attributed to two design choices. First, Sharpe-based universe selection effectively filtered high-variance and low-efficiency stocks, concentrating exposure on assets with consistent return-per-risk characteristics. Second, liquidity-constrained weighting, derived from volume and ADV-based adjustments, prevented over-allocation to thinly traded equities—reducing transaction costs and improving real-world tradability. The daily rebalancing mechanism further stabilized portfolio dynamics, maintaining a turnover rate of 44%—a balanced level between agility and cost control.

## 6. Conclusion and Future Work

This paper proposed a Sharpe-driven stock selection and liquidity-constrained portfolio optimization framework tailored to the Chinese equity market. By integrating risk-adjusted stock filtering, liquidity-aware mean–variance optimization, and drawdown-based risk management, the framework bridges the gap between theoretical portfolio design and practical market execution. Empirical backtests conducted on Shanghai Composite constituents from 2023 to 2025 show that the proposed method achieves a Sharpe ratio of 1.71, an annualized return of 25%, and a maximum drawdown of 8.2%, outperforming both classical benchmarks and contemporary machine-learning-based approaches. Compared with equal-weight, capitalization-ranked, and mean-variance models, our method delivers superior risk-adjusted returns while maintaining realistic trading conditions and moderate turnover. The findings provide robust evidence that liquidity-aware risk optimization and Sharpe-based selection can jointly enhance performance stability in emerging and partially efficient markets such as China. Beyond its empirical results, the framework offers practical implications for quantitative asset managers. It demonstrates that robust portfolio construction can be achieved by combining disciplined statistical screening with execution-aware constraints—without relying on complex, opaque models. Moreover, the adaptive rebalancing and drawdown control mechanisms ensure capital preservation under volatile market regimes, making the approach suitable for real-world deployment within both institutional and retail contexts.

Future research may extend this work along several directions. First, incorporating commodities such as gold and silver as a hedging layer could improve portfolio diversification and inflation resilience. Second, exploring alternative datasets—including financial news, market sentiment, fundamental indicators, or options-implied data—could enrich the stock-selection process. Third, alternative risk measures, such as the Sortino ratio or Conditional Value at Risk (CVaR), may offer more nuanced perspectives on downside protection. Fourth, integrating machine-learning-based feature extraction could refine Sharpe estimation and selection efficiency. Finally, extending the framework to multi-asset portfolios within the Chinese market—encompassing equities, bonds, and commodities—would provide a broader test of scalability and cross-asset adaptability. In summary, this study contributes both theoretically and practically to the field of quantitative portfolio optimization in emerging markets. The results confirm that combining Sharpe-driven selection with liquidity constraints yields a powerful and implementable framework for achieving consistent, risk-adjusted outperformance.